\documentclass[11pt,preprint2]{aastex}

\shorttitle{Matter in Spirals: I. Surface Photometry}
\shortauthors{Kassin et al.}

\begin{document}

\title{Dark and Baryonic Matter in Bright Spiral Galaxies:
I.Near-infrared and Optical Broadband Surface Photometry of 30
  Galaxies\altaffilmark{1}}

\shorttitle{Dark \& Baryonic Matter in Galaxies: Photometry}

\altaffiltext{1}{Based in part on observations obtained at the
Cerro Tololo Interamerican Observatory, operated by the Association of
Universities for Research in Astronomy, Incorporated, under a
cooperative agreement with the National Science Foundation.}

\author{Susan A. Kassin\altaffilmark{2,3},
Roelof S. de Jong\altaffilmark{4}, \&
Richard W. Pogge\altaffilmark{2}}

\altaffiltext{2}{Department of Astronomy, The Ohio State University,
140 W. 18th Ave., Columbus, OH 43210-1173}
\altaffiltext{3}{currently at: UCO/Lick Observatory, 
University of California, Santa Cruz, CA 95064; kassin@ucolick.org} 
\altaffiltext{4}{Space Telescope Science Institute, 3700 San Martin Dr., 
Baltimore, MD 21218}

\begin{abstract}
We present photometrically calibrated images and surface
photometry in the $B, V, R, J, H,$ and $K$-bands of 25,
and in the $g$, $r$, and $K$-bands
of 5 nearby bright ($B^o_T<12.5$ mag) spiral galaxies with
inclinations between $30$--$65$ degrees
spanning the Hubble Sequence from Sa to Scd.  Data are from The Ohio
State University Bright Spiral Galaxy Survey, the Two Micron All Sky
Survey, and the Sloan Digital Sky Survey Second Data Release.
Radial surface brightness profiles are extracted, and
integrated magnitudes are measured from the profiles.  Axis ratios,
position angles, and scale lengths are measured from the near-infrared images.
A 1-dimensional bulge/disk decomposition is performed
on the near-infrared images of galaxies with a non-negligible bulge component,
and an exponential disk is fit to the radial surface brightness profiles
of the remaining galaxies.
\end{abstract}


\section{Introduction}
The main uncertainty in the determination of the 
distribution of dark matter in galaxies from their rotation curves stems
from the poorly known stellar mass
distribution of galaxies \citep[e.g.,][]{verh}.  In the currently
fashionable picture of galaxy formation in a cold dark matter universe, 
both dark and luminous matter are expected to make a measurable 
contribution to the total mass in the inner parts of bright spiral galaxies.  
The relative contributions of luminous and dark matter have
so far been poorly constrained due to uncertainty in the mass scaling of the
stellar component, and the ensuing degeneracies 
between the stellar mass and dark matter contributions \citep[e.g.,][]{barn}.  
Recently, it has been shown that the uncertainty 
in the stellar component can be reduced by
the inclusion of multi-band photometry, preferably with at least one
near-infrared passband to abate the effects of dust on the
determination of the stellar populations present \citep{bdej}.  

The color versus mass-to-light ratio relations of stellar populations
described by \citet{bdej} are affected by dust attenuation.
It is known from detailed dust models (e.g., Disney, Davies, \& Phillipps 1989, 
Gordon et al.\ 2001, Witt, Thronson, \& Capuano 1992) that the effects of
dust on the surface brightness profiles and integrated magnitudes of galaxies
(extinction and scattering) are regulated by the geometry
of stars and dust in the galaxies.  Because this geometry is unknown,
we cannot know the exact effects of 
dust on the galaxies' surface brightness profiles and integrated magnitudes.
However, to first order, errors in the dust reddening
estimates are not expected to strongly affect the masses derived
from broadband colors of stellar populations.  The dust will systematically both
redden and extinguish galactic light, making a galaxy appear dimmer and 
to have a larger stellar mass-to-light ratio than it should.  These 
effects work in opposite directions, so that errors 
in extinction corrections lead only to small errors
in the derived stellar masses.  The effects of dust on the derivation of
stellar masses from photometry are discussed in depth in
a subsequent paper (hereafter Paper II; Kassin et al., accepted).

With a better determination of the stellar mass distribution,
one can place tighter constrains on the dark matter component of galaxies by
comparing the luminous components of a galaxy (stars and gas) to
total dynamical mass distributions derived from rotation curves.  
In this paper, we present optical and near-infrared photometry 
for a sample of galaxies that have rotation curves available in
the literature.   We use these data, along with surface brightness
profiles from \citet{verh}, the Sloan Digital Sky Survey Second 
Data Release \citet[SDSS DR2;][]{abaz}, and the Two Micron All Sky Survey
\citep[2MASS;][]{jar2,cutr,jar1} to determine the distributions of 
luminous and dark matter for 34 galaxies in Paper II.  
With this data in hand, we also investigate the angular 
momentum content of the 34 galaxies in \citet{kas3} (Paper\,III).

Historically, galaxies have been studied with broad-band photometry at
optical wavelengths, while more recent studies have extended into 
the near-infrared. Whereas optical bands are most sensitive to young 
Population I blue stars which account for only a small fraction of 
the total stellar mass of a galaxy, near-infrared bands give a much 
better census of the older stars which play a greater role in
determining the stellar mass of a galaxy.  Moreover, the extinction
due to dust is much less at near-infrared than at optical wavelengths. 
Extinction at the near-infrared {\it K}-band is only about $10 \% $ of 
that at the optical {\it B}-band \citep{mart}.  Numerous studies have focused 
on near-infrared photometry of local spiral galaxies 
\citep[e.g.,][]{dej1,hera,gava,knap,jar1}.
However, of these, only the \citet{verh} sample of both high and low 
surface brightness galaxies in Ursa Major includes
dynamical information.  The sample presented here encompasses 
non-interacting, large, bright,
spiral galaxies, for which most theories of galaxy formation and evolution 
make predictions, and which are useful for probing the spiral galaxy
Hubble Sequence.

This paper is organized as follows:
The details of our sample selection, observation, reduction,
and calibration are discussed in $\S 2$--4.  The calculation of 
physical parameters for each of the galaxies is discussed in $\S 5$.  
Our photometry and radial surface brightness profiles are
compared with those in the literature in $\S 6$.
Distances are calculated in $\S 7$, and 
a very brief summary of our data is presented in $\S 8$.  
Throughout this paper we
adopt a Hubble constant of H$_{\rm o}=70$ km s$^{-1}$ Mpc$^{-1}$.  When
distance-dependent quantities have been derived from the literature, we
have converted them using this value of $H_o$.

\section{Sample Selection}
The sample encompasses a total of 30 galaxies from The Ohio State University Bright
Spiral Galaxy Survey \citep[OSUBSGS;][]{eskr}, 2MASS,
and the SDSS DR2 which span the Hubble Sequence of spirals from Sa
to Scd for bright galaxies ($B_T^o <12.5$
mag). Table 1 lists all the galaxies in the sample, their
morphological class, $\mu_B= 25$ mag arcsec$^{-2}$ isophotal
diameter ($D_{25}$) from \citet{drc3} (hereafter, RC3),
$D_{25}$ measured from data in this paper,
heliocentric radial velocity ($V_{hel}$), and adopted distance.
Uncertainties in $D_{25}$ due to both errors in fits to the 
exponential disks and zero-point calibrations are typically $\sim 5$ arcsec. 
Parameters of the five galaxies
from the \citet{verh} sample that will be used in the subsequent analyses
are also listed.  Of the galaxies studied in this paper, 25 have
imaging in at least one band from the 2MASS, and 5 have optical
imaging from the SDSS DR2.  We use 2MASS $K$-band data to flux
calibrate the OSUBSGS near-infrared images, and in some cases, to
replace them.  For one galaxy, NGC\,3319, quality imaging at 
$K$ is not available, so we use a 2MASS $H$-band image to calibrate
its $H$-band image from the OSUBSGS.

The OSUBSGS is a sample of nearly 200 nearby bright spiral galaxies.
Galaxies in this survey were chosen from the RC3 to
have $1\leq \rm T_{RC3} \leq 7$ where T$_{\rm RC3}$ is the mean numerical
Hubble stage index as tabulated in the RC3, $B^o_T \leq 12$ mag,
$D_{25} \leq 6.5$ arcminutes where $D_{25}$ the apparent major isophotal
diameter measured at $\mu_B$=25 mag arcsec$^{-2}$, and $-80 < \delta
<+50$ degrees where $\delta$ the declination
(due to the pointing limits of the Cerro Tololo
Interamerican Observatory (CTIO) 1.5-meter and the
Lowell Observatory Perkins 1.8-meter telescopes).
We imposed a few additional criteria to select galaxies from the OSUBSGS:
galaxies were required (1) to be non-interacting, (2) within an optimal 
inclination range ($\sim30$--$\sim65$ degrees) in order to 
reduce the effect of dust extinction and reddening while still being able
to obtain accurate kinematical information, (3) to
have Galactic latitudes where the absorption due to our Galaxy is
quantified in \citet{schl}, (4) to have a photometric 
optical calibration in the OSUBSGS, (5) to have a photometric
near-infrared calibration in the OSUBSGS or the 2MASS, (6) and,
to have a rotation curve available from the literature.

Galaxies were selected from the SDSS DR2 and 2MASS surveys if
they had (1) sufficient quality imaging to create a high
signal-to-noise surface brightness profiles at $g, r$, and $K$,
(2) $B_T^o<12$ mag to satisfy the selection criterion of
the OSUBSGS, and satisfy the requirements for galaxies to be 
selected from the OSUBSGS.

\section{Observations and Reduction for the OSUBSGS Galaxies}
Data for the OSUBSGS were obtained during a large number of observing runs
with six telescopes of apertures between 0.9 and 1.8 meters during the
period 1993--2000.  Table 2 lists the telescope, instrument, and
detector used for each final image in our sample, along with the
date each image was taken. The observations were made by the
OSUBSGS team and their students, as well as by a professional observer
(Roberto Aviles) hired by the project at CTIO.
For details about the telescopes and instruments used 
and the manner in which the observations were taken, see \citet{eskr}.  



\section{Photometric Calibration}
\subsection{OSUBSGS \& 2MASS Galaxies}
OSUBSGS observing nights were judged to be photometric in three stages.  First,
at the telescope the observer made a decision based on the
weather and acquired standard star images if the night was apparently
photometric. Second, the standard star data were reduced and if it was
verified that the residuals between the observed and cataloged
standard star magnitudes did not change during the night was flagged
as possibly photometric. Finally, the
photometric zero-points for contiguous nights
were checked for consistency.  If the zero-points for a night were not
consistent with adjoining nights, and there was no change to the
observing set-up (i.e., instrument changes between nights)
that could account for the difference, then the night was judged
to be non-photometric, and images from those nights were excluded
from further consideration.  If the zero-points for a night were consistent 
with adjoining nights, and there was no change in the observing set-up,
then the night was judged to be photometric.

For the optical photometric calibration, equatorial standard star fields
from \citet{land} were observed at a range of airmasses to derive photometric
transformations onto the Kron-Cousins $BVR$ system.
Photometric zero-points, airmass terms, and $B-V$ color terms were
calculated for each night with the $\tt fitparams$ task in IRAF.
For each night, a photometric solution was fit to the standard
star observations which included both airmass and $B-V$ color terms.
Standard star observations which were outliers to this solution were
examined and generally were found to be problematic
due to factors such as cosmic ray contamination, bad pixels,
bad columns/rows, or because some stars were imaged near the edges
of the detectors.  The uncertainties given include those in the
photometric calibration, in the measurement of instrumental magnitudes
(which are dominated by sky variation), and that due to the fact
that a galaxy does not color-correct like a star (unless each
pixel of its image were individually color-corrected).

OSUBSGS $J, H,$ and $K$-band images were photometrically
calibrated with data from 2MASS.  An attempt was made 
to use standard stars from the list of
\citet{cart}, but due to ambiguities in the data we decided to
calibrate them against the 2MASS database instead.
To calibrate the 2MASS images, the zero-points listed
in the 2MASS image headers were applied to the images.

$J_{2MASS}$, $H_{2MASS}$, and $K_{S,2MASS}$ were transformed 
to $J_{CIT}$, $H_{CIT}$, and $K_{CIT}$ in order to calibrate the OSUBSGS 
near-infrared images that are on the CIT photometric system. 
The transformations used are the ones tabulated on the 2MASS
website\footnote[3]{http://www.astro.caltech.edu/$\sim$jmc/2mass/v3/transformations/} 
(where we leave out a color term that contributes as 0.001 mag):
\begin{eqnarray}
J_{CIT} = K_{S,2MASS} + (0.019\pm0.004)\\
J_{CIT} = K_{CIT} + \frac{(J-K_{S})_{2MASS} + 0.02}{1.068}\\
H_{CIT} = K_{CIT} + (H-K_{S})_{2MASS} - 0.034.
\end{eqnarray}

We calibrated the near-infrared OSUBSGS images to the 2MASS
images by comparing surface brightness profiles of galaxies
extracted from both surveys in the same manner.  We explain in detail
the extraction of these profiles in Paper\,II. 
Surface brightness profiles from the OSUBSGS were calibrated
to 2MASS profiles by finding the best-fit combination of a
sky determination and magnitude zero-point that
allows for the smallest difference between the profiles.
The OSUBSGS profiles generally have poorer sky
measurements, but they have higher signal-to-noise ratios
in their inner parts, and extend to larger galactocentric radii essential for
the subsequent analysis.  On average, the OSUBSGS infrared images have
limiting surface brightnesses of $\sim 2$ mag arcsec$^{-2}$ fainter than
those from 2MASS.  However, in some cases, the 2MASS images are
of better overall quality than the OSUBSGS images.  In those cases,
we adopt the 2MASS image for the final dataset. The calibration was done by plotting
the calibrated flux from a 2MASS profile against the instrumental counts
from an OSUBSGS profile.  A straight line was fit to these data.
The slope of this line determines the bootstrap zero-point
necessary to calibrate the OSUBSGS image, and its intercept determines
the sky offset multiplied by the zero-point.  This procedure works
well since, for a typical galaxy in the sample, the near-infrared 
color terms are small ($\sim 1\%$).

The photometric zero-point calibration of the 2MASS galaxies is
accurate to $\pm 2\%$--$3\%$ \citep{cutr}.  However, as warned in Appendix A
of \citet{jar2}, a small fraction of the galaxies may
be affected by high-frequency background variations, causing the
photometric error to increase.  This does not appear to be the case in
those galaxies presented here.  The sky variation
in our near-infrared images causes a further $\sim 2\%$ uncertainty in
the photometric calibration.  In total, the near-infrared calibrations
are uncertain to $\sim 4\%$. 

\subsection{SDSS DR2 Galaxies}
For the SDSS DR2 images, the zero-point calibration is accurate to
$\pm 2\%$ in both SDSS $r$ and SDSS $g-r$ \citep{lupt}; the sky variation
in our galaxy images causes a further $\sim 2\%$ uncertainty.  In total, the
zero-point calibration is uncertain by $\sim 4\%$.  For these SDSS DR2 galaxies,
we applied the zero-points and extinction terms as
given in the ``best TsField'' FITS table.

\section{Surface Brightness Profiles and Physical Properties of the Galaxies}
\subsection{Axis Ratios $\&$ Position Angles}
\label{sec:axispa}
In order to create surface brightness profiles, we first determined
the axis ratio and position angle of each galaxy.
(For galaxies with SDSS DR2 and 2MASS images,
we adopted the position angles and axis ratios used in the literature
for measuring their rotation curves.)  Axis ratios and position angles 
were measured from $H$-band images of the galaxies.  
The $H$-band was chosen to measure physical parameters,
because near-infrared wavelengths trace most of the stellar mass in galaxies,
and our $H$-band images generally have higher signal-to-noise than
our $J$ or $K$ images.  For each $H$-band image of a galaxy, ellipses were
fit with increasing semi-major axis from the galaxy's center.
This was done with the $\tt profile$ command in the
XVista\footnote[4]{XVista is based on Lick Observatory Vista and
maintained by a loose consortium of die-hard users at
http://ganymede.nmsu.edu/holtz/xvista.} package, which
uses a modification of \citet{kent}'s implementation of
the azimuthal Fourier moments technique as described by \citet{laue}.
The resulting plots of position angles and axis ratios of the
ellipses versus radius were examined by eye, and a position angle
and inclination were chosen for each galaxy at radii where the galaxy's disk was
exponential (usually between 2 and 3 scale lengths).  As a check on
the adopted parameters,
an ellipse with the chosen axis ratio and position angle was plotted over
the $H$-band image and visually inspected.  This procedure was
repeated until the parameters derived for each galaxy passed a visual
inspection.  The goal was to find the parameters that best follow the main
structure of the galaxies' stellar mass distributions.
The measured axis ratio is converted into an inclination
angle, $i$, under the assumption that the disks are intrinsically circular ($q=cos\,i$,
where $q$ is the ratio of minor and major axes).
The final position angles and inclinations adopted for the galaxies
are presented in columns 7 and 8 of Table 3.  The typical errors in the position angles and inclinations are both approximately $\pm 5$ degrees.

\subsection{Radial Surface Brightness Profiles}
\label{sec:sbprofiles}
To extract surface brightness profiles for each of our sample galaxies
we used the XVista command $\tt annulus$.
The SDSS DR2 galaxies were first aligned to the World Coordinate System so that
they were aligned with their respective 2MASS images.
The $\tt annulus$ command computes a radial surface brightness profile by
finding the median surface brightness per pixel in elliptical annuli of
increasing distance from the center of a galaxy.
We chose to calculate the median (instead of the mean)
surface brightness in order to avoid foreground
stars and effects such as bad pixels which tend to corrupt the average
statistic.  Ellipse parameters were pre-determined for each galaxy, as
discussed in \S\ref{sec:axispa}, and centers were defined as the pixel in the
nucleus with the highest surface brightness.

The resulting surface brightness profiles for each galaxy are presented 
in Figure 1, alongside their $B$ and $K$-band images 
($K_{S}$ for 2MASS images) to allow for comparison of surface
brightness profiles with morphological properties. The profiles
have been corrected for Galactic extinction using
Schlegel et al. (1998).
The galaxies NGC\,1090, NGC\,2841, and NGC\,3198
have SDSS DR2 images in which almost half of the
galaxy is not present, as can be observed in the $g$-band
images in Figure 1.  This is also the case, but to a much
lesser extent, for NGC\,3521.  We use the areas common to all passbands 
to create color profiles.

Total magnitudes and magnitudes within the $\mu_B=25$ mag arcsec$^{-2}$ isophotal radius
($R_{25}$) were measured by integrating the surface brightness profiles to
$R_{25}$, extrapolating the profiles with an exponential function when 
necessary.  The resulting magnitudes are listed in columns 4 and 5 of
Table 3 along with magnitudes from the literature in column 7
for comparison.  Note that \citet{verh}'s
$K^{\prime}$-band measurements have not been converted to $K_{CIT}$
since the difference between the two bands is small ($\sim
0.05$ mag for typical mean $J-H$ colors).  The seeing for each image 
is listed in column 3 of Table 3; the typical error in the seeing is $\pm 0.2 \arcsec$.


\subsection{Bulge/Disk Decompositions}
The radial surface brightness profiles in the $H$ or $K$-band for each
of the galaxies were decomposed into bulge and disk components
following \citet{dejb} and \citet{knap} (see also MacArthur, Courteau, 
\& Holtzman 2003).  The bulge component was fit using a generalized 
\citet{sers} profile of the form
\begin{equation}
\mu(\rm R) = \mu_e + 2.5 \rm b_n \left[ \left(R \over R_e
  \right)^{1/n} - 1 \right]
\end{equation}
where $n$ is the bulge shape parameters ($n=4$ for a deVaucouleurs
$r^{1/4}$ law, $n=1$ for an exponential disk), $R_{e}$ is the
effective radius, $\mu_{e}$ is the surface brightness at $R_{e}$,
and $b_{n}$ is a normalization factor depending on $n$ that
ensures that half of the integrated light is within $R_{e}$.  Only
$n$, $R_{e}$, and $\mu_{e}$ are free parameters.  The disk was
fit with a standard exponential surface brightness profile of the form
\begin{equation}
\mu(R) = \mu_0 + 1.086 (R/h)
\end{equation}
where $\mu_0$ is the central surface brightness of the disk and $h$ is
the disk scale length.

The results for the 5-parameter bulge/disk fits to our near-infrared
radial surface-brightness profiles are summarized in Table 4.  
The bulge-to-disk ratio $(B/D)$ was derived from integrating the best fit
bulge and disk profiles, and is listed in Table 4.
For 20 galaxies, the added bulge component did not
change the predicted stellar mass rotation curves beyond the uncertainties in our 
adopted mass-to-light ratios (see Paper\,II) compared
to the one derived for the disk alone.  We re-fit these galaxies solely with
an exponential disk.  Not surprisingly, these galaxies are
all among the latest Hubble types in this sample.

Note that the disk parameter measurements made here do not depend on the 
outer parts of the surface brightness profiles where uncertainties
in the sky determination have the most effect.  The outer
parts of the surface brightness profiles (where the signal-to-noise
drops below $\sim 1 \sigma$) were removed for the 
bulge/disk decompositions and exponential disk fits.
Because of this, the effect of uncertainties in the sky determination
also do not affect the bulge parameters.

\section{Comparison With Photometry and Radial Surface Brightness Profiles in the Literature}
In Figure 2, surface brightness profiles 
are compared with those from the literature and 
SDSS DR2.  We do not compare the near-infrared profiles
since they are calibrated to 2MASS.
We plot in Figure 2 the radial difference between the literature magnitude
and the magnitude derived in this paper.  In each of the panels of the figure, the bands of the 
surface brightness profiles compared are noted beneath the galaxy's NGC number:  
the bands of the profiles from this paper are written first, then those 
from the literature next.  Aside from the SDSS DR2 comparisons, those from  
the literature are from: \citet{ryd2}
for NGC\,157, \citet{math} for NGC\,908, NGC\,1241, NGC\,1385, NGC\,1559,
NGC\,1832, NGC\,2090, NGC\,2139, NGC\,7083, and NGC\,7606, 
\citet{hera} for NGC\,3726 and NGC\,4062,
\citet{ken6} for NGC\,4062, NGC\,7217, and NGC\,7606,
and \citet{buta} for NGC\,6300.  

\begin{figure}[!t]
\figurenum{2}
\includegraphics[height=3.6in, width=3in]{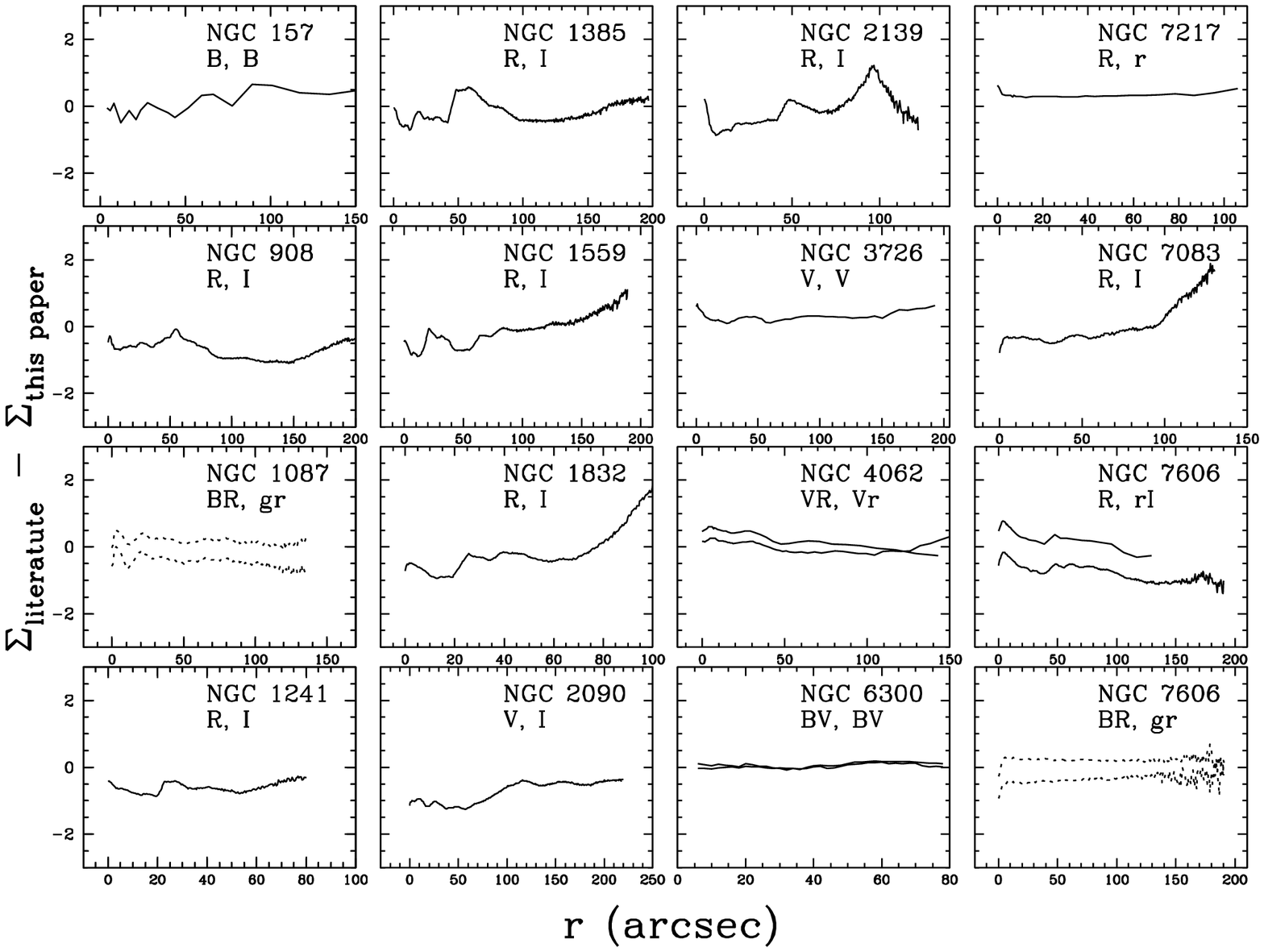}
\vspace{-0.8in}
\caption{{\small Surface brightness profiles from this paper 
are compared to those in the literature (solid lines) and the SDSS DR2 (short dashed lines).
Under the galaxy's NGC number, the band(s) of the surface brightness profile(s) 
plotted from this paper is(are) noted first, then the band(s) from the literature/SDSS DR2
are(are) noted after.}}
\end{figure}

For the SDSS DR2, we compare the $g$ and $r$-band (AB magnitudes) surface brightness
profiles with those for the $B$ and $R$-bands (Vega magnitudes) measured in this
paper.  There is an offset from zero for all the curves.
For NGC\,1087 and NGC\,7606, on average, the offsets for $g-B$ are 0.45 and
0.36 mag, and those for $r-R$ are 0.18 and 0.22, respectively.
This is consistent with the expected
colors of galaxies from the 1996 version of the Bruzual \& 
Charlot 2003 GISSEL models \citep{bruz} for exponentially declining star-formation
rates: $g-B \sim -0.38$ and
$r-R \sim 0.22$ for a typical galaxy color of $V-R = 0.5$.
There is not a large difference
between the sky subtraction of the SDSS profiles and those
presented in this paper since the sky values in both images were both
calculated with the same technique.

The difference between the $I$-band profiles measured by Mathewson et al.\ (1992) and
the $R$-band profiles from this paper is within the spread of $R-I$ colors
of spiral galaxies \citep{dej1}. The difference for $V$ compared with $I$
for NGC\,2090 is similarly within the spread of $V-I$ colors given in \citet{dej1}.
The Mathewson et al.\ (1992) profiles 
are consistent with those presented in this paper even though the 
methods that are used to compute them differ.  Those from this paper
are computed in a fixed ellipse determined from an outer isophote of an
$H$-band image, whereas those in Mathewson et al.\ (1992)
are computed in ellipses whose position angles and inclinations
are allowed to vary on the $I$-band images.  
A difference in profile extraction can also cause differences in the shapes of the 
surface brightness profiles, especially when dealing with bars and rings.
Also, the Mathewson et al.\ (1992) data have systematically higher sky values
as evidenced by the upturn at the ends of the curves in Figure 2.
Differences in sky subtraction are most
apparent in the outer parts of the profiles since it 
it there that the source
counts become comparable to the sky counts.  A difference in the
methods of sky determination may account for the differences in sky subtraction.
We calculate the
sky counts in our images by using the XVista command $\tt sky$ in boxes
inside the images where the contribution from the galaxies is small.
This command finds the sky background level under the assumption
that the most common pixel value in the chosen box is the sky value.
In particular, the $\tt sky$ routine calculates the mean of the pixel values in the
box, and builds a histogram of the values about the mean.  The region
of the peak pixel value is located in the histogram by fitting it with a 
parabola.  The center of this parabola
is taken as the sky value.  Mathewson et al.\ (1992) measured the sky by calculating the mode pixel value
in a 127 pixel wide boarder around each frame.  
Also, the regions where the sky value was calculated in Mathewson et al.\ (1992)
may have been contaminated by galaxy counts.

For comparisons with literature references other than SDSS DR2 and
Mathewson et al.\ (1992), the profiles are consistent.  For NGC\,157,
the profiles are consistent to within 0.05 mag, which is within
their uncertainties.  However, the sky measurements differ between
the two profiles.  Sky measurements were made by \citet{ryd2}
by using the modal peak of the histogram of data values within
20 pixels of the image edges, similar to the method used in this paper.
Therefore, we should not expect a difference between the sky measurements.
The integrated magnitude of NGC\,157 from \citet{ryd2} is also
fainter than the integrated magnitude given in this paper (which 
is consistent the RC3 measurement).
For NGC\,3726, we compare the $V$-band profiles 
and find a 0.30 mag difference on average.
This is consistent with the uncertainties: the uncertainty given by \citet{hera}
is 0.02 mag for the rms error due to the sky value uncertainty, the zero-point
uncertainty from \citet{hera} is estimated to be within 0.1--0.3 mag, and the 
zero-point uncertainty for the profile presented in this paper is 0.06 mag.
For NGC\,4062, the average difference from the literature is 0.06 mag,
which is within uncertainties.
For NGC\,6300, the $B$ and $V$ profiles are compared with profiles of the same
bands from the literature, and the differences are found to be 0.05 and 0.06 mag, 
respectively, consistent with zero-point uncertainties.
The galaxies NGC\,4062, NGC\,7217, and NGC\,7606 have average offsets between
the $R$-band profiles given here the and $r$-band profiles given in \citet{kent} of 
0.34, 0.33, and 0.36 mag, respectively.  These differences are consistent with the 
expected colors of typical galaxies in this paper (see paragraph on SDSS comparison)
if zero-point uncertainties are taken into account.

In Table 3, we list integrated magnitudes from the literature
for many of the galaxies in our sample.  
We plot the difference between total
magnitudes from the RC3 and those presented in this paper
in Figure 3a,b for the $B$ and $V$-bands, respectively.  For the $B$-band, 
we find a mean difference of 0.13 mag with a $\sigma$ of 0.33 mag. 
For the $V$-band, there are only 4
galaxies with a measurement in the RC3.  For this band, there is no mean difference between
the magnitudes presented here and in the RC3, 
but there is a $\sigma$ of 0.29 mag.  The zero-point differences 
of these comparisons are within
the expected errors, but the high $\sigma$ values are somewhat
disturbing.  Those galaxies with large differences between zero-points 
are NGC\,289 (differences of 0.67 mag for $B$ and 0.55 mag for $V$),
NGC\,1241 (0.89 mag for $B$), and NGC\,2280 (0.90 for $B$).  For NGC\,289, 
the $V$-band magnitude given for data in this paper, while not consistent with
the RC3, is consistent with that of \citet{wals}.
Similarly, for NGC\,1241, the $R$-band surface brightness profile 
given in this paper is 
consistent with the $I$-band profile from Mathewson et al. (1992).  
And, for NGC\,2280, the $B$-band magnitude is consistent with 
that of \citet{laub}.

Other integrated magnitude measurements from the literature
are for NGC\,157 from \citet{ryd2} (discussed above), 
NGC\,3726, NGC\,4062, NGC\,4651, and NGC\,7606 from \citet{hera}, 
and NGC\,6300 from \citet{buta}.  For 
NGC\,4062, NGC\,4651, and NGC\,7606, the magnitudes are consistent.
The differences between the integrated magnitudes of NGC\,6300
given here and in the literature are likely due to
the myriad foreground stars contaminating its image.  Integrated magnitudes given in the literature
for NGC\,6300 are calculated by removing a few field stars from the 
image and then using aperture photometry.
\citet{buta} used photographic plates and they mention that ``a wide range
of apertures was used, but because of the lack of a conspicuous nucleus
and the large number of foreground stars, it was not possible to obtain the maximum
range achievable with the photographic equipment and telescopes used for the
observing.''  Integrated magnitudes are derived from the surface brightness
profiles of NGC\,6300 which were calculated with a median statistic.  This allows us to 
avoid the problem of subtracting the myriad foreground stars in the image of NGC\,6300. 
This difference can explain why the surface brightness profiles of NGC\,6300
are consistent with those in the literature, while the integrated magnitudes are not.

\begin{figure}[!t]
\figurenum{3}
\includegraphics[height=3in, width=3.6in]{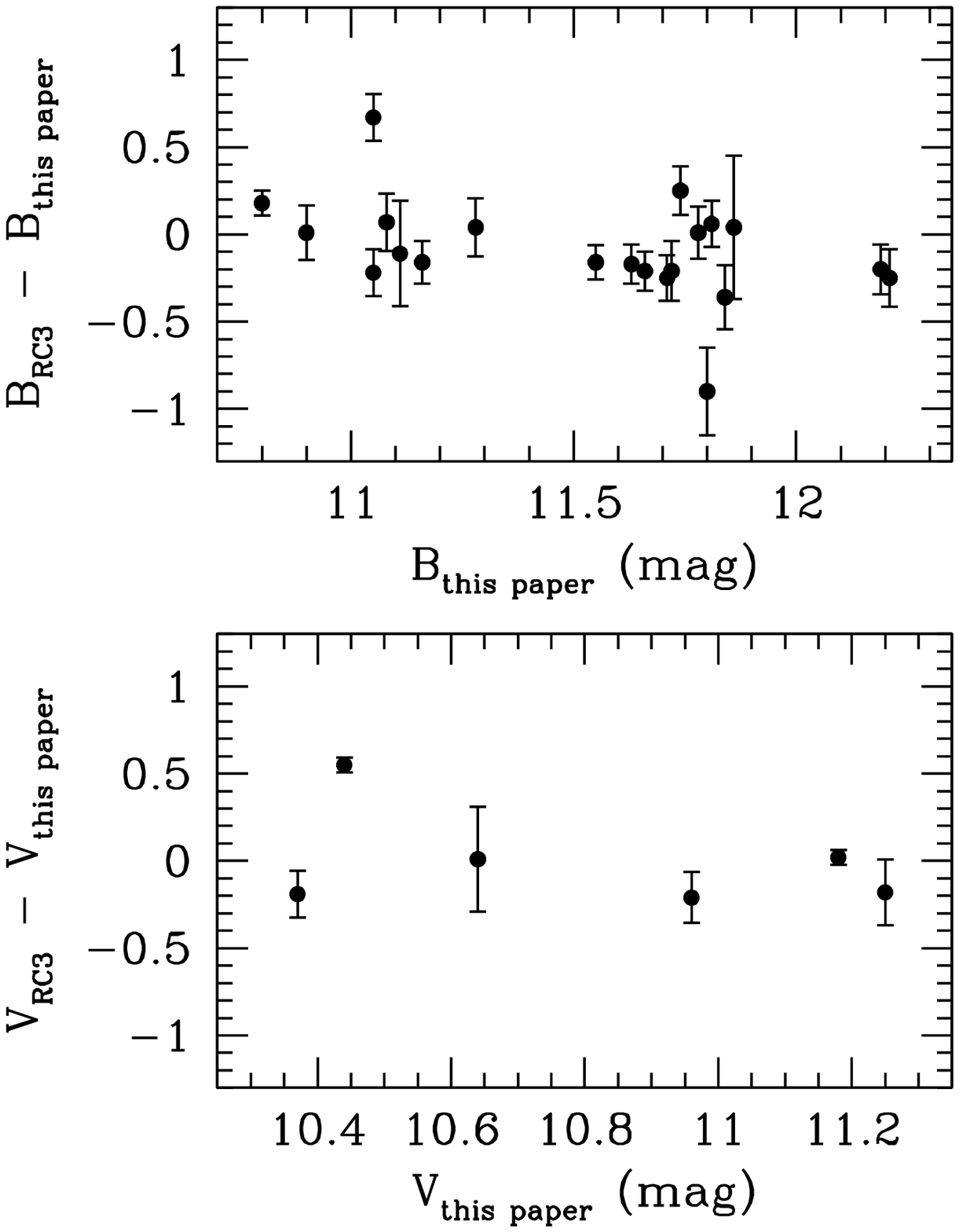}
\caption{{\small We plot the difference between the total integrated RC3 magnitudes and the integrated 
magnitudes from this paper versus magnitudes from this paper.  In the upper panel we compare the $B$-band
magnitudes, and in the lower panel we compare the $V$-band magnitudes.}}
\end{figure}

\section{Distances}
Table 1 lists the distances for each of the galaxies in Megaparsecs (Mpc).
These were calculated under the assumption of Hubble flow, after
correction for Virgocentric infall, following the formalism
of \citet{aaro}.
Four galaxies were found to have triple-valued solutions for their
distances in the Virgocentric infall solution.  For these galaxies,
NGC\,4062, NGC\,4651, and NGC\,4698,
a distance was adopted based on $H$-band
Tully-Fisher distances.  

In addition, 4 galaxies in the sample have distances measured by
Hubble Space Telescope (HST) observations of Cepheid variable stars: NGC\,2090
\citep{phel}, NGC\,2841 \citep{macr}, NGC\,3198 \citep{kels}, and
NGC\,3319 \citep{saka}. In all 4 cases, we have adopted
these Cepheid distances, which are listed in Table 1.  For those
galaxies from \citet{verh}, we adopt the HST Key Project distance 
to the Ursa Major cluster of 20.7 Mpc \citep{sak1}.

\section{Summary}
Photometrically calibrated surface brightness profiles,
magnitudes, and physical parameters are presented for a a sample of 
31 nearby bright spiral galaxies for which dynamical information is 
available in the literature.  


\acknowledgements
SAK would like to acknowledge financial support from The Space
Telescope Science Institute Director's Discretionary Research Fund (DDRF).
We thank the referee, P. Grosb$\o$l, for his thorough comments that helped
improve the final version of this paper.
We are grateful to Jay Frogel for initiating the OSUBSGS and to Don Terndrup
for help with the optical calibrations used in this paper.

We thank the CTIO TAC for generous allocation of time for the OSU 
Galaxy Survey and the many people over the years who helped collect 
these observations.  Funding for the OSU Bright Spiral Galaxy Survey was 
provided by grants from The National Science Foundation 
(grants AST-9217716 and AST-9617006), with additional funding by The 
Ohio State University.  

This paper makes use of data from both the Sloan Digital Sky Survey and the
Two Micron All Sky Survey.  The Two Micron All Sky Survey is a joint 
project of the University of Massachusetts and the Infrared Processing 
and Analysis 
Center/California Institute of Technology, funded by the National 
Aeronautics and Space Administration and the National Science Foundation.
Funding for the creation and distribution of
the SDSS Archive has been provided by the Alfred P. Sloan Foundation,
the Participating Institutions, the National Aeronautics and Space
Administration, the National Science Foundation, the U.S. Department
of Energy, the Japanese Monbukagakusho, and the Max Planck
Society. The SDSS Web site is http://www.sdss.org/.
The SDSS is managed by the Astrophysical Research Consortium (ARC) for
the Participating Institutions. The Participating Institutions are The
University of Chicago, Fermilab, the Institute for Advanced Study, the
Japan Participation Group, The Johns Hopkins University, Los Alamos
National Laboratory, the Max-Planck-Institute for Astronomy (MPIA),
the Max-Planck-Institute for Astrophysics (MPA), New Mexico State
University, University of Pittsburgh, Princeton University, the United
States Naval Observatory, and the University of Washington.

This research has made use of the NASA/IPAC 
Extragalactic Database (NED) that is operated by the Jet Propulsion 
Laboratory, California Institute of Technology, 
under contract with the National Aeronautics and Space Administration.
This research has also made use of NASA's Astrophysics Data System.

{\it Facilities:} \facility{Perkins (OSIRIS,IFPS)}, 
\facility{CTIO:0.9m (CFCCD)}, \facility{CTIO:1.5m (CIRIM, CFCCD)}, 
\facility{Hiltner (MIS)}, \facility{McGraw-Hill (MIS)}

\clearpage
\begin{figure*}[right]
\figurenum{1a}
\includegraphics[scale=0.75]{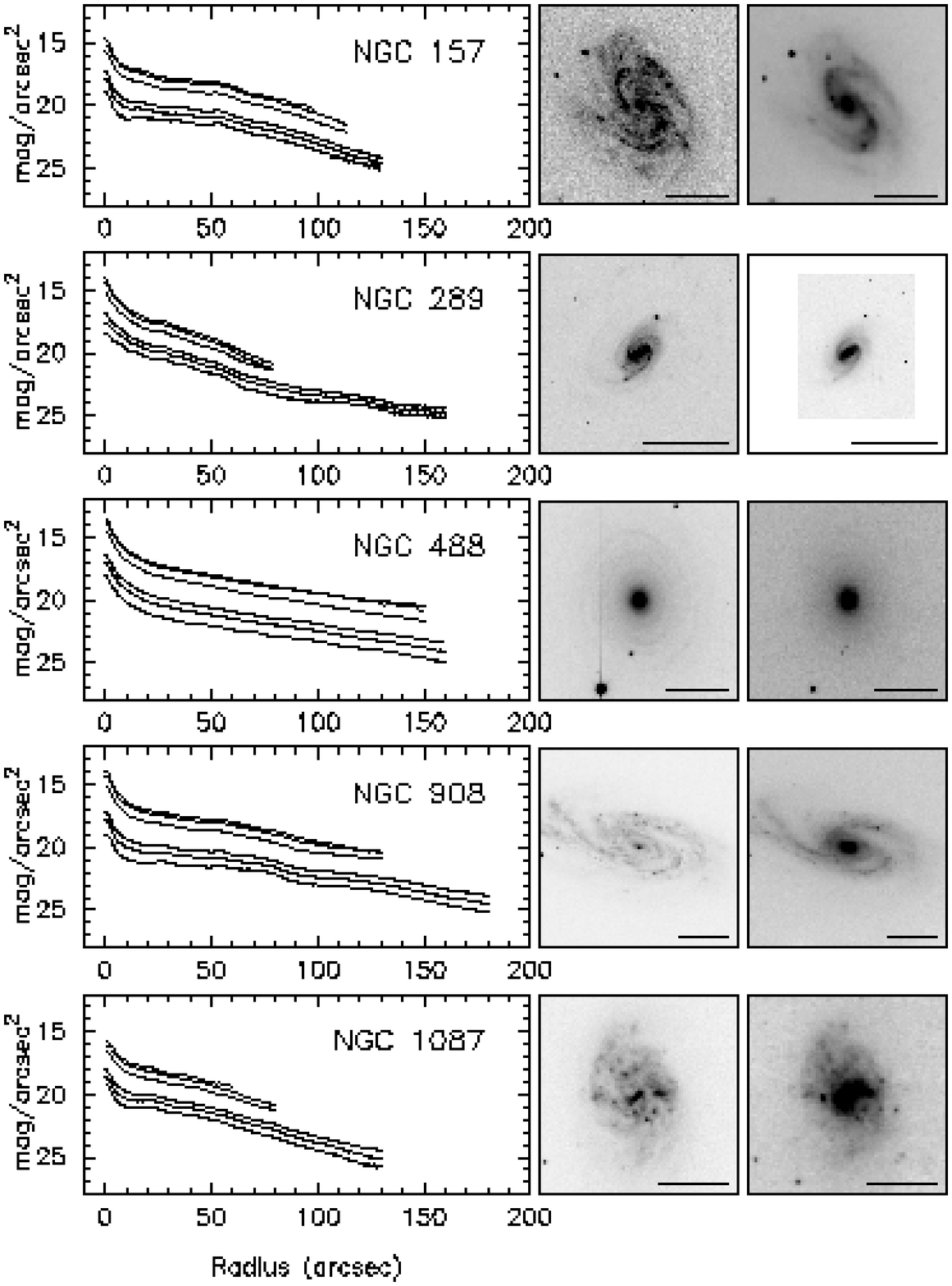}
\caption{{\small Observed surface brightness profiles and $B$ and $K$-band images
for the sample galaxies.  In this and subsequent figures, surface
brightness curves run B-to-K bottom-to-top; see Table~\ref{tab:tab2ch2}
for the observed bands.  The $B$ image is on the left, $K$ on the right,
and the scale bar indicates 60\arcsec.  Orientation is North=Up,
East=Left.  Degraded resolution for astro-ph; please see journal article
or send an email for a version at a higher resolution.}}
\end{figure*}

\clearpage
\begin{figure*}
\figurenum{1b}
\includegraphics[scale=0.8]{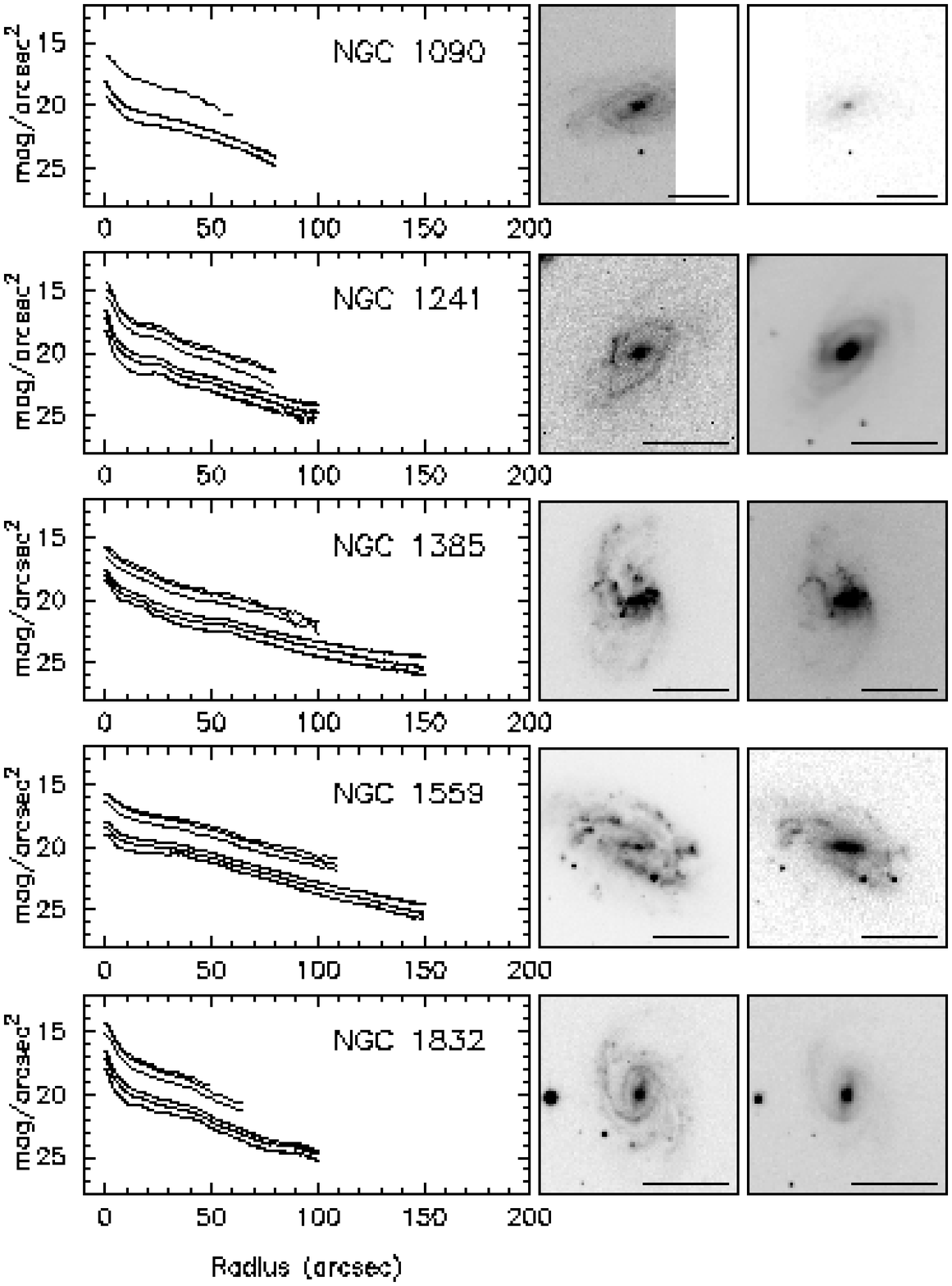}
\caption{{\small See Figure 1a.}}
\end{figure*}

\begin{figure*}
\figurenum{1c}
\includegraphics[scale=0.8]{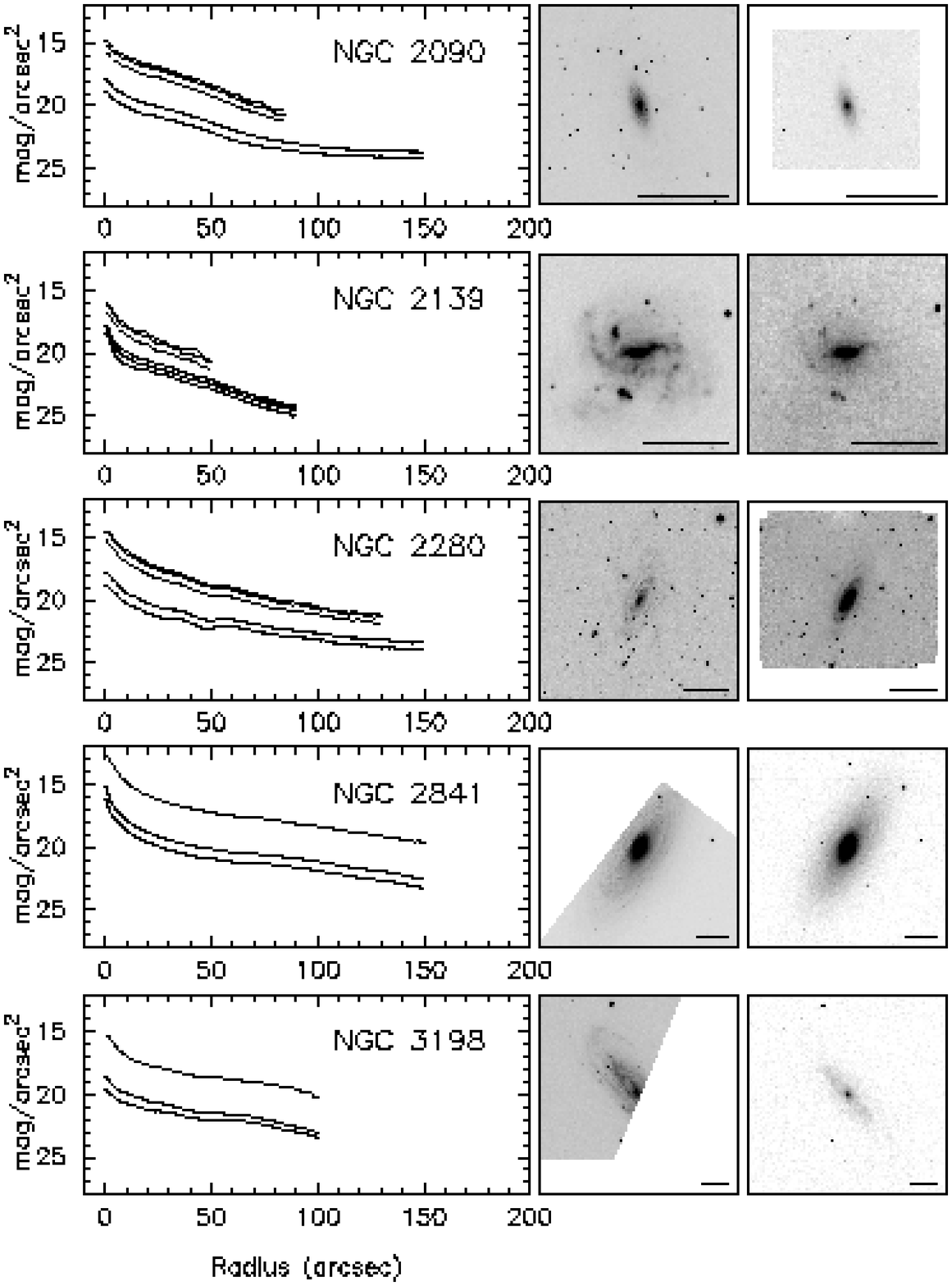}
\caption{{\small See Figure 1a.}}
\end{figure*}

\begin{figure*}
\figurenum{1d}
\includegraphics[scale=0.8]{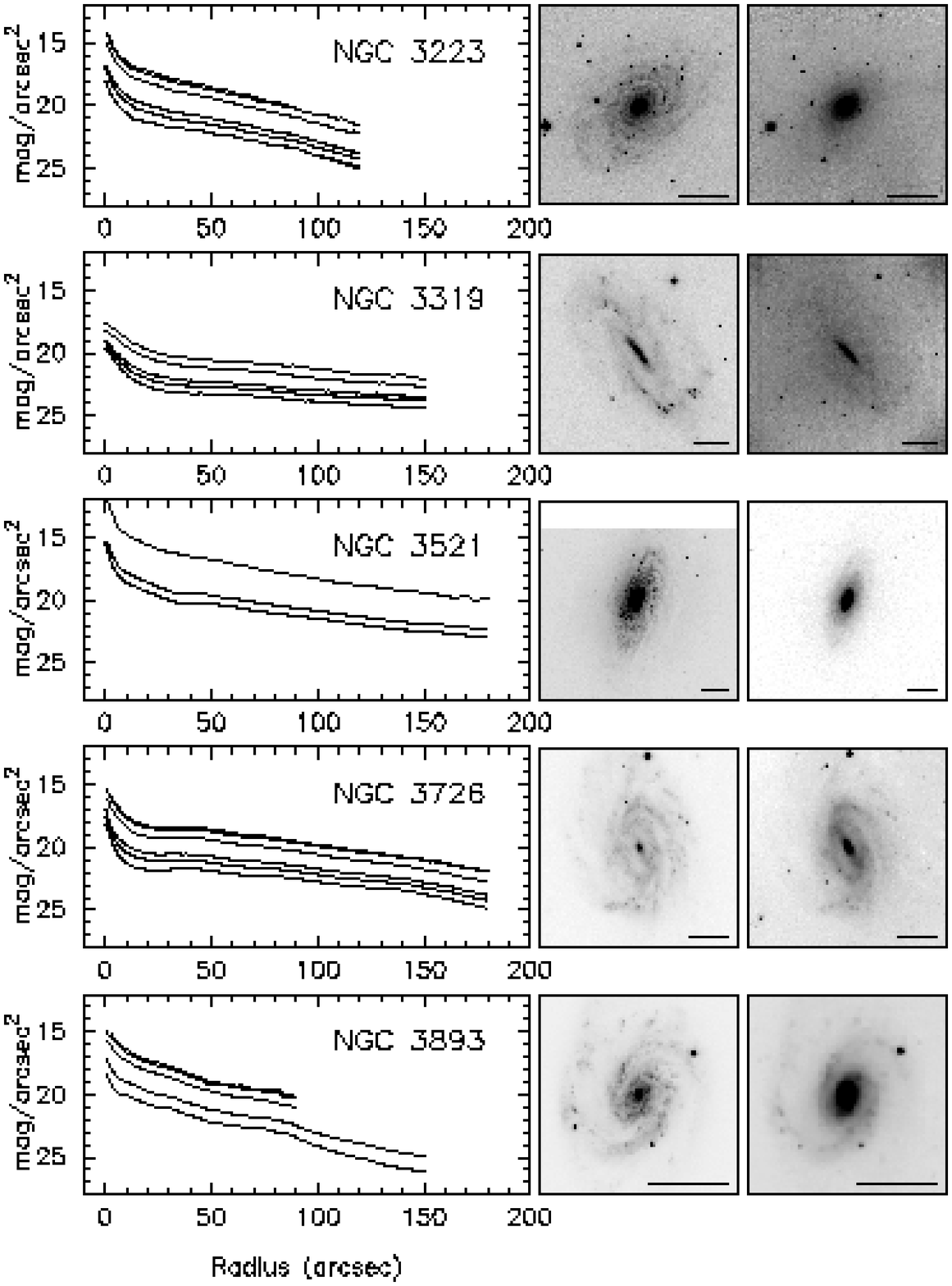}
\caption{{\small See Figure 1a.}}
\end{figure*}

\begin{figure*}
\figurenum{1e}
\includegraphics[scale=0.8]{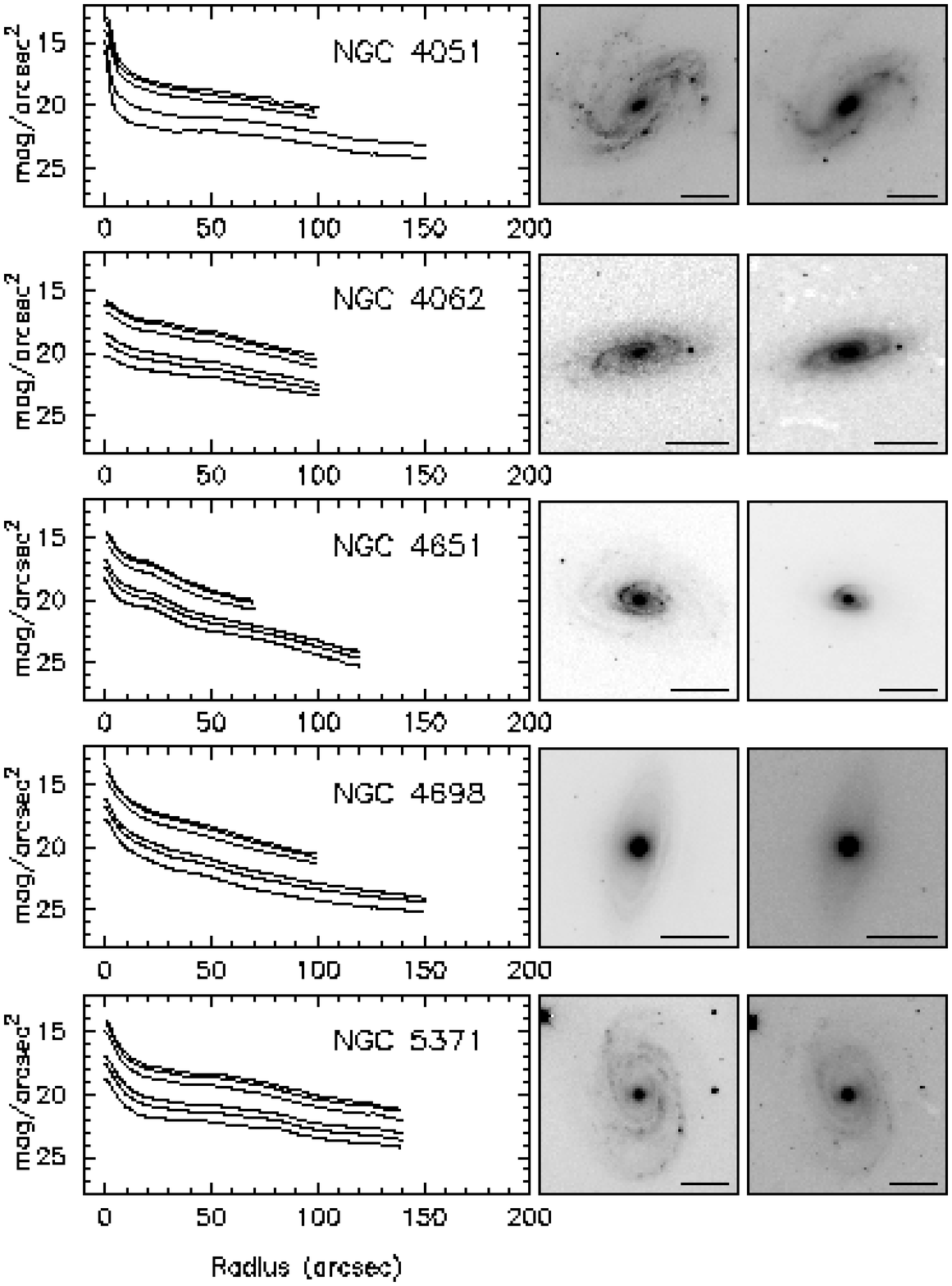}
\caption{{\small See Figure 1a.}}
\end{figure*}

\begin{figure*}
\figurenum{1f}
\includegraphics[scale=0.8]{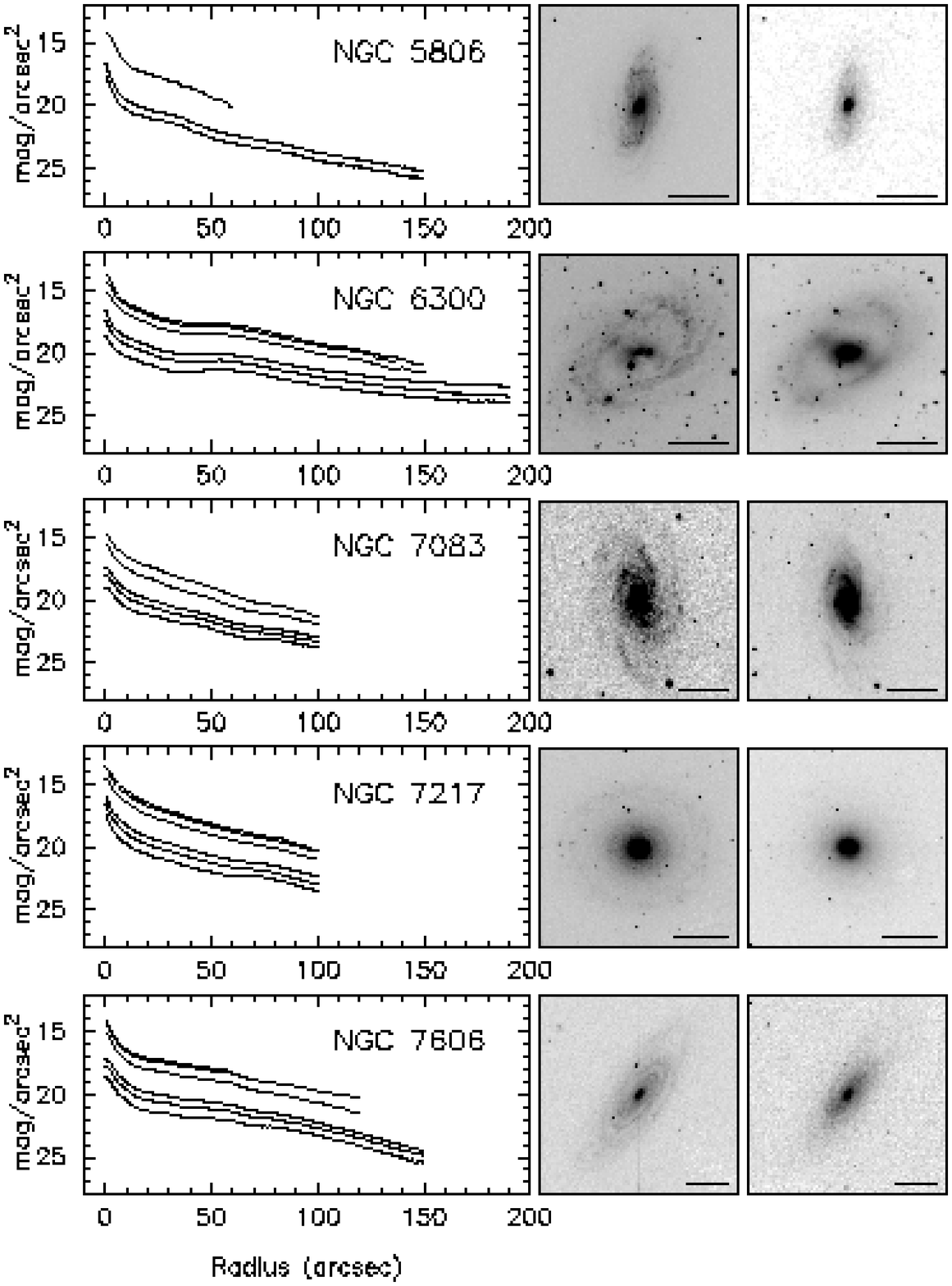}
\caption{{\small See Figure 1a.}}
\end{figure*}

\begin{deluxetable}{llcccccc}
\tabletypesize{\small}
\tablecaption{\label{tab:tab1ch2}Basic parameters of the sample galaxies.}
\tablehead{
\colhead{}&
\colhead{}&
\colhead{}&
\colhead{}&
\colhead{$D_{25}$ (RC3)} &
\colhead{$D_{25}$ (this paper)\tablenotemark{a}} &
\colhead{$V_{hel}$}&
\colhead{Distance}\\
\colhead{}&
\colhead{}&
\colhead{Galaxy}&
\colhead{RC3 Type} &
\colhead{($\arcsec$)}&
\colhead{($\arcsec$)}&
\colhead{(km s$^{-1}$)}&
\colhead{(Mpc)}}
\startdata
& &\object{NGC 157} &SAB(rs)bc   & 250 &253 & 1668   &21.6\\
& &\object{NGC 289} &SAB(rs)bc   & 308 &238 & 1628   &19.9\\
& &\object{NGC 488} &SA(r)b      & 315 &332 & 2272   &31.0\\
& &\object{NGC 908} &SA(s)c      & 362 &350 & 1498   &18.9\\
& &\object{NGC 1087} &SAB(rs)c    & 223 &219 & 1519   &20.7\\
& &\object{NGC 1090}\tablenotemark{b} &SB(rs)bc    & 239 &230 &2758    &37.9\\
& &\object{NGC 1241} &SB(rs)b     & 169 &177 & 4052   &55.5\\
& &\object{NGC 1385} &SB(s)cd     & 203 &208 & 1493   &19.3\\
& &\object{NGC 1559} &SB(s)cd     & 208 &249 & 1292   &15.8\\
& &\object{NGC 1832} &SB(r)bc     & 154 &184 & 1937   &27.3\\
& &\object{NGC 2090} &SA(rs)b     & 294 &291 &\nodata &12.3\tablenotemark{e}\\
& &\object{NGC 2139} &SAB(rs)cd   & 158 &171 & 1843   &26.2\\
& &\object{NGC 2280} &SA(s)cd     & 379 &378 & 1906   &27.5\\
& &\object{NGC 2841}\tablenotemark{b} &SA(r)b      & 488 &483 &\nodata &14.1\tablenotemark{e}\\
& &\object{NGC 3198}\tablenotemark{b} &SB(rs)c     & 511 &298 &\nodata &14.5\tablenotemark{e}\\
& &\object{NGC 3223} &SA(r)bc     & 244 &246 & 2895   &43.6\\
& &\object{NGC 3319} &SB(rs)cd    & 370 &364  &\nodata      &14.3\tablenotemark{e}\\
& &\object{NGC 3521}\tablenotemark{b} &SAB(rs)bc   & 658 &486 & 805   &9.4\\
& &\object{NGC 3726} &SAB(r)c     & 370 &398 &  866   &21.7\\
& &\object{NGC 3893}\tablenotemark{c} &SAB(rs)c    & 268 &236 &  967   &23.6\\
& &NGC 3949\tablenotemark{d} &SA(s)bc     & 173 &176 &\nodata &20.7\\
& &NGC 3953\tablenotemark{d} &SB(r)bc     & 415 &424 &\nodata &20.7\\
& &NGC 3992\tablenotemark{d} &SB(rs)bc    & 455 &532 &\nodata &20.7\\
& &\object{NGC 4051} &SAB(rs)bc   & 315 &306 &\nodata & 15.2\\
& &\object{NGC 4062} &SA(s)c      & 244 &273 &  769   &10.8\tablenotemark{f}\\
& &NGC 4138\tablenotemark{d} &SA(r)0+     &144 &154  &\nodata &20.7\\
& &\object{NGC 4651} &SA(rs)c     & 239 &229 &  805   &18.3\tablenotemark{f}\\
& &\object{NGC 4698} &SA(s)ab     & 239 &241 & 1002   &19.1\tablenotemark{f}\\
& &\object{NGC 5371} &SAB(rs)bc   & 262 &320 & 2553   &45.1\\
& &\object{NGC 5806}\tablenotemark{c} &SAB(s)b     & 185 &193 & 1359   &26.4\\
& &\object{NGC 6300} &SB(rs)b     & 268 &430 & 1110   &14.5\\
& &\object{NGC 7083} &SAB(rs)c    & 233 &276 & 3109   &40.5\\
& &\object{NGC 7217} &(R)SA(r)ab  & 233 &224 &  952   &15.9\\
& &\object{NGC 7606} &SA(s)b      & 322 &279 & 2233   &29.4\\
\enddata
\tablenotetext{a}{Typical uncertainties are $\sim 5$ arcsec.}
\tablenotetext{b}{All imaging is from SDSS DR2 \& 2MASS}
\tablenotetext{c}{All imaging is from \citet{verh} \& 2MASS}
\tablenotetext{d}{All imaging is from \citet{verh}.  Distance is taken as the
HST Key Project distance to the Ursa Major cluster (Sakai et al.\ 2000), so
no heliocentric velocity is listed.}
\tablenotetext{e}{Distance measured from Hubble Space
  Telescope observations of Cepheid variable stars, so no heliocentric velocity 
is listed.}
\tablenotetext{f}{Galaxies which have a triple-valued solution for the
  Virgo infall calculation.  The chosen solution is the one that is the
  closest to the calculated Tully-Fisher distance (given $W_r$
  and H$_{-0.5}$ from Tully (1988).}
\end{deluxetable}

\begin{deluxetable}{ccclcllll}
\tabletypesize{\small}
\tablecaption{\label{tab:tab2ch2}Observational Details for the OSUBSGS}
\tablehead{
\colhead{}&
\colhead{}&
\colhead{}&
\colhead{Galaxy}&
\colhead{Bands}&
\colhead{Date}&
\colhead{Telescope}&
\colhead{Camera}&
\colhead{Detector}
}
\startdata
& & &NGC\,157&BVR&1995 Oct 29&CTIO 0.9m&CFCCD&Tek1K\#2\\
& & & &JH&1995 Oct 08&Perkins 1.8m&OSIRIS&NICMOS3\\
& & & &K&2MASS&\nodata&\nodata&\nodata\\
& & &NGC\,289&BVR&1995 Oct 26&CTIO 0.9m&CFCCD&Tek1K\#2\\
& & & &JK&1996 Oct 01&CTIO 1.5m&CIRIM&NICMOS3\\
& & & &H&1996 Oct 02&CTIO 1.5m&CIRIM&NICMOS3\\
& & &NGC\,488&BVR&1994 Oct 11&Perkins 1.8m&IFPS&NCCD\\
& & & &JHK&1995 Oct 18&Perkins 1.8m&OSIRIS&NICMOS3\\
& & &NGC\,908&BVR&1995 Oct 26&CTIO 0.9m&CFCCD&Tek1K\#2\\
& & & &JHK&1996 Sep 30&CTIO 1.5m&CIRIM&NICMOS3\\
& & &NGC\,1087&BVR&1993 Sep 17&Perkins 1.8m&IFPS&NCCD\\
& & & &J&1995 Oct 15&Perkins 1.8m&OSIRIS&NICMOS3\\
& & & &H&1995 Oct 06&Perkins 1.8m&OSIRIS&NICMOS3\\
& & & &K&2MASS&\nodata&\nodata&\nodata\\
& & &NGC\,1241&BVR&1995 Oct 29&CTIO 0.9m&CFCCD&Tek1K\#2\\
& & & &JH&1995 Oct 10&Perkins 1.8m&OSIRIS&NICMOS3\\
& & & &K&2MASS&\nodata&\nodata&\nodata\\
& & &NGC\,1385&BVR&1995 Oct 27&CTIO 0.9m&CFCCD&Tek1K\#2\\
& & & &JHK&2MASS&\nodata&\nodata&\nodata\\
& & &NGC\,1559&BVR&1995 Oct 26&CTIO 0.9m&CFCCD&Tek1K\#2\\
& & & &JHK&1995 Mar 10&CTIO 1.5 m&CIRIM&NICMOS3\\
& & &NGC\,1832&BVR&1994 Nov 01&CTIO 0.9m&CFCCD&Tek2K\#3\\
& & & &JHK&2MASS&\nodata&\nodata&\nodata\\
& & &NGC\,2090&BV&1995 Mar 07&CTIO 1.5m&CFCCD&Tek1K\#2\\
& & & &JHK&1995 Mar 08&CTIO 1.5m&CIRIM&NICMOS3\\
& & &NGC\,2139&BVR&1994 Apr 09&CTIO 0.9m&CFCCD&Tek1K\#2\\
& & & &JHK&2MASS&\nodata&\nodata&\nodata\\
& & &NGC\,2280&BV&1995 Mar 08&CTIO 1.5m&CFCCD&Tek1K\#2\\
& & & &R&\nodata&\nodata&\nodata&\nodata\\
& & & &JH&1995 Mar 09&CTIO 1.5m&CIRIM&NICMOS3\\
& & & &K&1995 Mar 15&CTIO 1.5m&CIRIM&NICMOS3\\
& & &NGC\,3223&BVR&1994 Apr 07&CTIO 0.9m&CFCCD&Tek1K\#2\\
& & & &JHK&1994 Feb 25&CTIO 1.5m&OSIRIS&NICMOS3\\
& & &NGC\,3319&BVR&2000 Apr 22&MDM 1.3m&MIS&Echelle\\
& & & &JH&1995 Apr 25&Perkins 1.8m&OSIRIS&NICMOS3\\
& & & &K&\nodata&\nodata&\nodata&\nodata\\
& & &NGC\,3726&BVR&2000 Apr 26&MDM 1.3m&MIS&Echelle\\
& & & &JH&1996 Mar 08&Perkins 1.8m&OSIRIS&NICMOS3\\
& & & &K&1997 Mar 15&Perkins 1.8m&OSIRIS&NICMOS3\\
& & &NGC\,4051&BR&Verheijen&\nodata&\nodata&\nodata\\
& & & &V&\nodata&\nodata&\nodata&\nodata\\
& & & &JH&1996 Mar 31&Perkins 1.8m&OSIRIS&NICMOS3\\
& & & &K&2MASS&\nodata&\nodata&\nodata\\
& & &NGC\,4062&BVR&1996 Feb 15&Perkins 1.8m&IFPS&NCCD\\
& & & &J&2MASS&\nodata&\nodata&\nodata\\
& & & &HK&1995 Mar 27&Perkins 1.8m&OSIRIS&NICMOS3\\
& & &NGC\,4651&BVR&1997 Mar 11&USNO&CCD&TI800\\
& & & &JHK&2MASS&\nodata&\nodata&\nodata\\
& & &NGC\,4698&BVR&1998 Mar 24&MDM 2.4m&MIS&Templeton\\
& & & &JH&1996 Apr 29&Perkins 1.8m&OSIRIS&NICMOS3\\
& & & &K&1997 Apr 07&Perkins 1.8m&OSIRIS&NICMOS3\\
& & &NGC\,5371&BVR&1996 Feb 15&Perkins 1.8m&IFPS&NCCD\\
& & & &JHK&1995 Mar 23&Perkins 1.8m&OSIRIS&NICMOS3\\
& & &NGC\,6300&BVR&1996 Apr 12&CTIO 0.9m&CFCCD&Tek2K\#3\\
& & & &JHK&2MASS&\nodata&\nodata&\nodata\\
& & &NGC\,7083&BVR&1995 Oct 26&CTIO 0.9m&CFCCD&Tek1K\#2\\
& & & &JK&1996 Sep 26&CTIO 1.5m&CIRIM&NICMOS3\\
& & &NGC\,7217&B&1994 Oct 10&Perkins 1.8m&IFPS&NCCD\\
& & & &VR&1993 Sep 18&Perkins 1.8m&IFPS&NCCD\\
& & & &JH&1995 Oct 18&Perkins 1.8m&OSIRIS&NICMOS3\\
& & & &K&2MASS&\nodata&\nodata&\nodata\\
& & &NGC\,7606&BVR&1994 Oct 11&Perkins 1.8m&IFPS&NCCD\\
& & & &JHK&1994 Oct 25&CTIO 1.5m&CIRIM&NICMOS3
\enddata
\end{deluxetable}

\begin{deluxetable}{ccccccccc}
\tabletypesize{\small}
\tablecaption{Measured Galaxy Parameters}
\tablehead{
\colhead{} &
\colhead{} &
\colhead{Seeing} &
\multicolumn{3}{c}{Magnitude} &
\colhead{$PA$} &
\colhead{$i$}\\
\cline{4-6}
\colhead{Galaxy} &
\colhead{Band} &
\colhead{($\arcsec$)} &
\colhead{total\tablenotemark{a}} &
\colhead{uncertainty} &
\colhead{Literature (total)} &
\colhead{(\arcdeg)} &
\colhead{(\arcdeg)}
}
\startdata
NGC 157  &B &1.3 &11.16 &0.03 &11.29$\pm$0.03\tablenotemark{b}, 11.00$\pm$0.12\tablenotemark{c} &43  &45.6 \\
         &V &1.1 &10.54 &0.03 &\nodata & & \\
         &R &1.0 &10.02 &0.03 &\nodata & & \\
         &J &2.6 &8.48  &\nodata &\nodata & & \\
         &H &2.6 &7.81  &\nodata &\nodata & & \\
         &K &2.9 &7.59  &\nodata &\nodata & & \\
NGC 289  &B &2.0 &11.05 &0.03 &11.72$\pm$0.13\tablenotemark{c} &148 &49.1 \\
         &V &1.8 &10.44 &0.03 &10.38$\pm$0.07\tablenotemark{d},10.99$\pm$0.03\tablenotemark{c} & & \\
         &R &1.4 &10.01  &0.03 &\nodata & & \\
         &J &1.9 &8.81  &\nodata &\nodata & & \\
         &H &1.5 &8.24  &\nodata &\nodata & & \\
         &K &1.6 &7.98  &\nodata &\nodata & & \\
NGC 488  &B &1.6 &11.08 &0.10 &11.15$\pm$0.13\tablenotemark{c} &10  &47.2 \\
         &V &1.5 &10.17 &0.04 &\nodata & & \\
         &R &1.8 &9.57  &0.07 &\nodata & & \\
         &J &2.4 &7.86  &\nodata &\nodata & & \\
         &H &2.6 &7.07  &\nodata &\nodata & & \\
         &K &2.4 &6.90  &\nodata &\nodata & & \\
NGC 908  &B &1.6 &11.05 &0.03 &10.83$\pm$0.13\tablenotemark{c} &80  &59.3 \\
         &V &1.1 &10.37 &0.03 &10.18$\pm$0.13\tablenotemark{c} & & \\
         &R &1.2 &9.81  &0.03 &\nodata & & \\
         &J &1.9 &8.17  &\nodata &\nodata & & \\
         &H &2.0 &7.54  &\nodata &\nodata & & \\
         &K &2.0 &7.33  &\nodata &\nodata & & \\
NGC 1087 &B &1.9 &11.71\tablenotemark{e} &0.05\tablenotemark{e} &11.46$\pm$0.12\tablenotemark{c} &12  &50.4 \\
         &V &2.1 &11.11\tablenotemark{e} &0.05\tablenotemark{e} &\nodata & & \\
         &R &1.7 &10.63\tablenotemark{e} &0.05\tablenotemark{e} &\nodata & & \\
         &J &2.4 &9.29  &\nodata &\nodata & & \\
         &H &2.6 &8.78  &\nodata &\nodata & & \\
         &K &1.9 &8.47  &\nodata &\nodata & & \\
NGC 1090 &g &1.6 &12.44\tablenotemark{f} &\nodata &\nodata &98  &56.6 \\
         &r &1.6 &11.64\tablenotemark{f} &\nodata &\nodata & &  \\
         &K &3.3 &9.24  &\nodata &\nodata & &  \\
NGC 1241 &B &3.1 &12.88 &0.03 &11.99$\pm$0.13\tablenotemark{c}  &128 &51.7 \\
         &V &3.6 &12.03 &0.03 &\nodata & & \\
         &R &2.5 &11.41 &0.03 &\nodata & & \\
         &J &2.6 &9.74  &\nodata &\nodata & & \\
         &H &2.6 &8.94  &\nodata &\nodata & & \\
         &K &1.9 &8.66  &\nodata &\nodata & & \\
NGC 1385 &B &1.2 &11.66\tablenotemark{e} &0.05 &11.45$\pm$0.10\tablenotemark{c} &170 &47.2 \\
         &V &1.0 &11.11\tablenotemark{e} &0.05 &\nodata & & \\
         &R &1.2 &10.63\tablenotemark{e} &0.05 &\nodata && & \\
         &J &2.0 &9.25 &\nodata  &\nodata & & \\
         &H &1.9 &8.66 &\nodata  &\nodata & & \\
         &K &1.9 &8.40 &\nodata  &\nodata & & \\
NGC 1559 &B &2.1 &11.11 &0.03 &11.00$\pm$0.30\tablenotemark{c} &66  &57.3 \\
         &V &1.8 &10.64 &0.03 &10.65$\pm$0.30\tablenotemark{c} & & \\
         &R &1.6 &10.21 &0.03 &\nodata & & \\
         &J &1.9 &8.85 &\nodata  &\nodata & & \\
         &H &2.0 &8.27 &\nodata  &\nodata & & \\
         &K &2.0 &8.05 &\nodata  &\nodata & & \\
NGC 1832 &B &1.6 &12.21 &0.10 &11.96$\pm$0.13\tablenotemark{c} &20  &48.7 \\
         &V &1.3 &11.53 &0.04 &\nodata & & \\
         &R &1.3 &11.01 &0.07 &\nodata & & \\
         &J &2.0 &9.33 &\nodata &\nodata & & \\
         &H &1.9 &8.67 &\nodata &\nodata & & \\
         &K &1.9 &8.41 &\nodata &\nodata & & \\
NGC 2090 &B &1.5 &11.74 &0.05 &11.99$\pm$0.13\tablenotemark{c} &14  &63.3 \\
         &V &1.4 &10.94 &0.04 &\nodata & & \\
         &R &\nodata &\nodata &\nodata &\nodata & & \\
         &J &2.0 &8.91 &\nodata &\nodata & & \\
         &H &2.0 &8.27 &\nodata &\nodata & & \\
         &K &1.9 &8.02 &\nodata &\nodata & & \\
NGC 2139 &B &2.0 &12.19 &0.06 &11.99$\pm$0.13\tablenotemark{c} &80  &39.2 \\
         &V &2.2 &11.73 &0.05 &\nodata & & \\
         &R &2.0 &11.34 &0.06 &\nodata & & \\
         &J &1.8 &10.19 &\nodata &\nodata & & \\
         &H &1.9 &9.58  &\nodata &\nodata & & \\
         &K &1.9 &9.33  &\nodata &\nodata & & \\
NGC 2280 &B &1.2 &11.80 &0.15 &10.90$\pm$0.20\tablenotemark{c}, 11.13$\pm$0.09\tablenotemark{x} &157 &66.4 \\
         &V &1.2 &11.03 &0.14 &\nodata & & \\
         &R &\nodata     &\nodata &\nodata &\nodata & & \\
         &J &2.0 &9.00 &\nodata  &\nodata & & \\
         &H &2.0 &8.37 &\nodata  &\nodata & & \\
         &K &2.0 &8.16 &\nodata  &\nodata & & \\
NGC 2841 &g &1.2 &9.79\tablenotemark{f} &\nodata   &\nodata &148 &56.63  \\
         &r &1.0 &9.03\tablenotemark{f} &\nodata   &\nodata &  &  \\
         &K &3.1 &6.13 &\nodata   &\nodata &  &  \\
NGC 3198 &g &1.2 &11.47\tablenotemark{f} &\nodata   &\nodata &30   &60  \\
         &r &1.4 &10.91\tablenotemark{f} &\nodata   &\nodata &  &  \\
         &K &3.5 &7.95  &\nodata   &\nodata &  &  \\
NGC 3223 &B &1.2 &11.78 &0.05 &11.79$\pm$0.14\tablenotemark{c}  &134 &40.5 \\
         &V &1.1 &10.95 &0.04 &\nodata & & \\
         &R &1.2 &10.36 &0.04 &\nodata & & \\
         &J &1.9 &8.53  &\nodata &\nodata & & \\
         &H &1.9 &7.82  &\nodata &\nodata & & \\
         &K &1.9 &7.60  &\nodata &\nodata & & \\
NGC 3319 &B &1.3 &11.84\tablenotemark{e} &0.07 &11.48$\pm$0.17\tablenotemark{c} &40  &60.0 \\
         &V &1.3 &11.25\tablenotemark{e} &0.06 &11.07$\pm$0.18\tablenotemark{c} & & \\
         &R &1.4 &10.94\tablenotemark{e} &0.06 &\nodata & & \\
         &J &2.4 &9.96 &\nodata  &\nodata & & \\
         &H &2.4 &9.36 &\nodata  &\nodata & & \\
         &K &\nodata &\nodata   &\nodata &\nodata & & \\
NGC 3521 &g &0.9 &9.36\tablenotemark{f} &\nodata    &\nodata &163 &53.13  \\
         &r &0.8 &8.65\tablenotemark{f} &\nodata    &\nodata &  &  \\
         &K &3.0 &7.97 &\nodata    &\nodata &  &  \\
NGC 3726 &B &1.9 &10.90 &0.14 &10.91$\pm$0.07\tablenotemark{c}  &9   &54.6 \\
         &V &1.5 &10.30 &0.06 &10.62 $\pm$0.02,0.1--0.3\tablenotemark{g}  & & \\
         &R &1.1 &9.81  &0.09 &9.97 & & \\
         &J &2.4 &8.63  &\nodata &\nodata & & \\
         &H &2.6 &7.94  &\nodata &\nodata & & \\
         &K &2.6 &7.77  &\nodata &\nodata & & \\
NGC 3893 &B &\nodata &11.38 &0.05 &11.16$\pm$0.15\tablenotemark{c} &172 &48 \\
         &V &\nodata &\nodata &\nodata  &\nodata & & \\
         &R &\nodata &10.28 &0.05 &\nodata & & \\
         &J &2.4 &8.72 &\nodata  &\nodata & & \\
         &H &2.4 &7.98 &\nodata  &\nodata & & \\
         &K &2.0 &7.83 &\nodata  &\nodata & & \\
NGC 3949 &B &\nodata &11.65 &0.05 &\nodata &117   &52  \\
         &R &\nodata &10.77 &0.05 &\nodata &  &  \\
         &K &\nodata &8.47 &\nodata &\nodata &  &  \\
NGC 3953 &B &\nodata &11.02 &0.05 &\nodata &13   &60  \\
         &R &\nodata &9.69 &0.05 &\nodata  &  &  \\
         &K &\nodata &7.00 &\nodata &\nodata &  &  \\
NGC 3992 &B &\nodata &10.66 &0.05 &\nodata &68   &56  \\
         &R &\nodata &9.45 &0.05 &\nodata  &  &  \\
         &K &\nodata &7.03 &\nodata &\nodata &  &  \\
NGC 4051 &B &\nodata  &11.28 &0.05 &\nodata &131   &49  \\
         &V &\nodata  &\nodata &\nodata &\nodata &  &  \\
         &R &\nodata &10.15 &0.05 &\nodata &  &  \\
         &J &2.4 &8.91 &\nodata &\nodata & & \\
         &H &2.4 &8.36 &\nodata &\nodata & & \\
         &K &2.6 &8.00 &\nodata &\nodata & & \\
NGC 4062 &B &2.6 &11.86 &0.09 &11.9$\pm$0.40\tablenotemark{c} &97 &63.9 \\
         &V &1.3 &11.18 &0.04 &11.20$\pm$0.01,$\lesssim 0.05$\tablenotemark{g} & & \\
         &R &1.3 &10.62 &0.04 &\nodata & & \\
         &J &2.5 &9.00  &\nodata &\nodata & & \\
         &H &2.6 &8.37  &\nodata &\nodata & & \\
         &K &2.6 &8.13  &\nodata &\nodata & & \\
NGC 4138 &B &\nodata &12.34 &0.05  &\nodata &151   &51.0  \\
         &R &\nodata &10.79 &0.05  &\nodata &  &  \\
         &K &\nodata &8.22 &\nodata  &\nodata &  &  \\
NGC 4651 &B &2.0 &11.55 &0.06 &11.39$\pm$0.08\tablenotemark{c} &82 &47.9 \\
         &V &1.9 &10.81 &0.04 &10.78 $\pm$0.03,$\lesssim 0.05$\tablenotemark{g} & & \\
         &R &1.9 &10.30 &0.04 &\nodata & & \\
         &J &2.6 &8.80 &\nodata  &\nodata & & \\
         &H &2.4 &8.16 &\nodata  &\nodata & & \\
         &K &2.4 &7.98 &\nodata  &\nodata & & \\
NGC 4698 &B &1.0 &11.63\tablenotemark{e} &0.06 &11.46$\pm$0.08\tablenotemark{c} & &168 &51.3 \\
         &V &1.0 &10.67\tablenotemark{e} &0.04 &\nodata & & \\
         &R &0.9 &10.14\tablenotemark{e} &0.04 &\nodata & & \\
         &J &2.7 &8.38  &\nodata &\nodata & & \\
         &H &2.6 &7.77  &\nodata &\nodata & & \\
         &K &2.4 &7.54  &\nodata &\nodata & & \\
NGC 5371 &B &2.1 &11.28 &0.09 &11.32$\pm$0.14\tablenotemark{c}  &16  &49.8 \\
         &V &2.0 &10.55 &0.04 &\nodata & & \\
         &R &2.0 &9.99  &0.04 &\nodata & & \\
         &J &2.4 &8.45 &\nodata  &\nodata & & \\
         &H &2.4 &7.81 &\nodata  &\nodata & & \\
         &K &2.4 &7.56 &\nodata  &\nodata & & \\
NGC 5806 &g &1.2 &12.10   &\nodata &\nodata &170 &58  \\
         &r &0.8 &11.37   &\nodata &\nodata &  &  \\
         &K &2.7 &8.46   &\nodata &\nodata  &  &  \\
NGC 6300 &B &1.1 &10.80 &0.05 &10.98$\pm$0.05\tablenotemark{c},10.98  $\pm$0.02\tablenotemark{i}  &123 &51.7 \\
         &V &1.1 &10.00  &0.03 &\nodata & & \\
         &R &0.9 &9.33  &0.03 &\nodata & & \\
         &J &1.9 &7.79  &\nodata &\nodata & & \\
         &H &1.9 &7.13  &\nodata &\nodata & & \\
         &K &1.9 &6.89  &\nodata &\nodata & & \\
NGC 7083 &B &2.4 &11.81 &0.03 &11.87$\pm$0.13\tablenotemark{c} &10  &58.0 \\
         &V &1.0 &11.19 &0.03 &\nodata & & \\
         &R &1.1 &10.70 &0.03 &\nodata & & \\
         &J &2.0 &9.20  &\nodata &\nodata & & \\
         &H &\nodata &\nodata &\nodata &\nodata & & \\
         &K &1.9 &8.28  &\nodata &\nodata & & \\
NGC 7217 &B &0.9 &11.26 &0.10 &\nodata &90 &29.5 \\
         &V &0.9 &10.40 &0.05 &\nodata & & \\
         &R &0.9 &9.74  &0.07 &\nodata & & \\
         &J &2.4 &7.88 &\nodata &\nodata & & \\
         &H &2.4 &7.16 &\nodata &\nodata & & \\
         &K &3.6 &6.93 &\nodata &\nodata & & \\
NGC 7606 &B &2.0 &11.72 &0.10 &11.51$\pm$0.14\tablenotemark{c} &146 &63.9 \\
         &V &1.9 &10.96 &0.04 &10.75$\pm$0.14\tablenotemark{c},10.98  $\pm$0.01,$\lesssim 0.05$\tablenotemark{g} & & \\
         &R &2.0 &10.41 &0.07 &\nodata & & \\
         &J &2.0 &8.61  &\nodata &\nodata & & \\
         &H &2.0 &7.79  &\nodata &\nodata & & \\
         &K &2.0 &7.67  &\nodata &\nodata & & \\
\enddata
\tablenotetext{a}{Uncorrected for Galactic extinction.  Uncertainties in the near-infrared
photometry are taken to be $\sim 4\%$.}  
\tablenotetext{b}{From Ryder et al.\ 1998; not extinction corrected.}
\tablenotetext{c}{From the RC3; not extinction corrected.}
\tablenotetext{d}{From Walsh et al.\ 1997; not extinction corrected.}
\tablenotetext{e}{Based on a secondary calibration obtained from a
  short ``snapshot'' image taken on a photometric night}
\tablenotetext{f}{For SDSS DR2 images, the zero-point calibration is accurate to 
$\pm 2\%$ in $r$ and $g-r$ (Lupton et al.\ 2001); the sky variation in our galaxy
images causes a further $\sim 2\%$ uncertainty.  In total, the zero-point calibration
is uncertain by $\sim 4 \%$.}
\tablenotetext{g}{From H\'{e}raudeau \& Simien 1996; not extinction corrected. The first error given is the
rms error due to uncertainty in the sky value.  The second error is the estimated accuracy 
in the zero-point.}
\tablenotetext{h}{No stars on image to accurately measure seeing.}
\tablenotetext{i}{From Buta 1987; not extinction corrected.}
\tablenotetext{x}{From Lauberts \& Valentijn 1989; not extinction corrected.}
\end{deluxetable}

\begin{deluxetable}{ccccccccc}
\tabletypesize{\small}
\tablecaption{\label{tab:tab2ch3}Bulge/Disk Parameters for $K$-band Images}
\tablehead{
\colhead{}
&\colhead{}
&\colhead{}
&\colhead{}
&\multicolumn{3}{c}{Bulge}
&\multicolumn{2}{c}{Disk}\\
\cline{5-7}
\cline{8-9}
\colhead{}
&\colhead{}
&\colhead{}
&\colhead{}
&\colhead{}
&\colhead{$R_e$}
&\colhead{$\mu_e$}
&\colhead{$h$}
&\colhead{$\mu_0$}\\
\colhead{}
&\colhead{}
&\colhead{Galaxy}
&\colhead{$B/D$}
&\colhead{$n$}
&\colhead{(arcsec)}
&\colhead{(mag/arcsec$^2$)}
&\colhead{(arcsec)}
&\colhead{(mag/arcsec$^2$)}
}
\startdata    
& & NGC 157\tablenotemark{a}  &\nodata    &\nodata    &\nodata &\nodata           &27  &16.3  \\
& & NGC 289                   &0.10    &1.10    &4  &15.7          &18  &15.9  \\
& & NGC 488                   &0.20    &2.20    &9  &16.4          &38  &16.6  \\
& & NGC 908\tablenotemark{a}  &\nodata    &\nodata    &\nodata &\nodata           &34  &16.3  \\
& & NGC 1087\tablenotemark{a} &\nodata    &\nodata    &\nodata &\nodata           &24  &16.8  \\
& & NGC 1090\tablenotemark{a} &\nodata    &\nodata    &\nodata &\nodata           &21  &17.1  \\
& & NGC 1241                  &0.20    &1.30    &4  &16.0          &17  &16.8 \\
& & NGC 1385\tablenotemark{a} &\nodata    &\nodata    &\nodata &\nodata           &26  &17.2  \\
& & NGC 1559                  &0.02    &1.20    &3  &17.1          &25  &16.4  \\
& & NGC 1832                  &0.15    &1.20    &3  &14.9          &15  &16.0  \\
& & NGC 2090\tablenotemark{a} &\nodata    &\nodata    &\nodata &\nodata           &17  &15.3  \\
& & NGC 2139\tablenotemark{a} &\nodata    &\nodata    &\nodata &\nodata           &17  &17.3  \\
& & NGC 2280                  &0.35    &1.30    &12  &16.9          &27  &16.7  \\
& & NGC 2841                  &0.19    &1.10    &9   &15.0          &30  &15.6  \\
& & NGC 3198\tablenotemark{a} &\nodata    &\nodata    &\nodata &\nodata           &39  &17.2  \\
& & NGC 3223\tablenotemark{a} &\nodata    &\nodata    &\nodata &\nodata           &25  &16.6  \\
& & NGC 3319\tablenotemark{a} &\nodata    &\nodata    &\nodata &\nodata           &72  &19.8  \\
& & NGC 3521                  &0.13    &1.40    &7   &14.7          &33  &15.1  \\
& & NGC 3726\tablenotemark{a} &\nodata    &\nodata    &\nodata &\nodata           &45  &17.4  \\
& & NGC 3893                  &0.29    &1.50    &14  &17.5          &28  &16.9  \\
& & NGC 3949\tablenotemark{a} &\nodata    &\nodata    &\nodata &\nodata           &16  &16.1  \\
& & NGC 3953\tablenotemark{a} &\nodata    &\nodata    &\nodata &\nodata           &39  &16.6  \\
& & NGC 3992\tablenotemark{a} &\nodata    &\nodata    &\nodata &\nodata           &56  &17.3  \\
& & NGC 4051                  &0.18    &3.70    &3   &14.7          &38  &17.4  \\
& & NGC 4062\tablenotemark{a} &\nodata    &\nodata    &\nodata &\nodata           &27  &16.3  \\
& & NGC 4138\tablenotemark{a} &\nodata    &\nodata    &\nodata &\nodata           &14  &15.6   \\
& & NGC 4651\tablenotemark{a} &\nodata    &\nodata    &\nodata &\nodata           &16  &15.7  \\
& & NGC 4698                  &0.28    &3.10    &7   &16.3          &24  &16.2  \\
& & NGC 5371                  &0.11    &1.30    &5   &15.7          &37  &17.0  \\
& & NGC 5806                  &0.20    &1.40    &4   &15.3          &19  &16.3  \\
& & NGC 6300                  &0.09    &1.20    &6   &15.9          &30  &16.2  \\
& & NGC 7083                  &0.06    &1.50    &3   &15.7          &16  &15.8  \\
& & NGC 7217                  &\nodata    &\nodata    &\nodata &\nodata           &24  &15.7  \\
& & NGC 7606                  &\nodata    &\nodata    &\nodata &\nodata           &31  &16.3  \\
\enddata
\tablenotetext{a}{For galaxies with a negligible bulge component we
  fit an exponential disk, and hence we only list the central
  surface brightnesses and scalelengths of the disks.}
\end{deluxetable}

\end{document}